\documentclass{article}
\usepackage{graphicx}
\begin{document}
\begin{flushright}
\small{IISc-CTS/13/00}
\end{flushright}
\begin{center}
\Large{A CALCULATION OF HIGGS MASS IN THE STANDARD MODEL}
\end{center}
%\singlespacing
\vspace{0.5in}
\begin{center}
J.Pasupathy

Centre for Theoretical Studies

Indian Institute of Science

Bangalore-560012 India.

email: jpcts@cts.iisc.ernet.in
\vspace{2in}

\large{\bf Abstract}
\end{center} 
 
  The assumption that the ratio of the Higgs self-coupling to
  the square of its yukawa coupling to the top is (almost)
  independent of the renormalization scale fixes the Higgs 
  mass within narrow limits  at $m_H = 160 GeV$ using only
  the values of  gauge couplings and top mass .

\newpage

     The suggestion that the mass of the Higgs may be symbiotically
 related to the mass of the top quark arose from analogy with 
 BCS like theories which predict the existence of a scalar
bound state at nearly twice the mass of the quasi particle in
the weak-coupling  limit [1]. Indeed top-condensate models[2]
given certain approximations do lead to a close relation
between the top and the Higgs mass. 
  On the other hand it was pointed out long ago point by Cabibbo
et.al.[3] that the assumption  that no new physics appears until
energies typical of Grand unified theories, combined with the
 use of the renormalisation group [RG] equations at the one loop 
level leads to both an upper bound and a lower bound on the 
Higgs  mass . As explained by them , these  bounds
come about due to the sensitive dependence of the solution of the RG
equations on the  initial value of the Higgs self coupling $\lambda (t)$
 at some low  energy scale. These studies were further extended by 
Beg et.al[4] , who considered energy scales up to the Landau pole ,
 and by Lindner[5] who extended it to lower energies [6]. 

  At the present time all details of the RG equations are known
 except for the integration constant of the equation for $\lambda (t)$.
 The question then arises whether there is some
  way to determine this constant which we  take to be the 
  value of $\lambda (t)$ ,  at the Z -mass , in terms of
  other known parameters of the standard model.  

 I now make the assumption that $\lambda (t)$  and the square of 
 the Higgs coupling to the top , $g_t^2(t)= G(t)$  have the same 
 t-dependence . Here  t is defined as
\begin{equation}
	t= \frac{1}{2} log(\frac{\mu^2}{m_Z^2})
\end{equation}                               
 with $\mu$ as the renormalization scale . Implementing this over 
 a modest range of values is sufficient to fix the value of
 $\lambda(0)$ within narrow limits and hence the Higgs mass.

  Consider the RG equations [7] for G(t) and $\lambda (t)$ 
\begin{equation}\label{two}
\frac{dG(t)}{dt}=\frac{1}{8\pi^2} \frac{9}{2} G(t)\{G(t)-\frac{17}{54} g_1^2(t)-\frac{1}{2}g_2^2(t)-\frac{16}{9}g_3^2(t)\}
\end{equation}

$$\frac{d\lambda}{dt}=\frac{1}{8\pi^2}.6\{\lambda^2(t)+\lambda(t)[G(t)-\frac{1}{4}g_1^2(t)-\frac{3}{4}g_2^2(t)]-$$
\begin{equation}
G^2(t)+\frac{1}{16}g_1^4(t)+\frac{1}{8}g_1^2(t)g_2^2(t)+\frac{3}{16}g_2^4(t)\}
\end{equation}

  Here $g_1 , g_2 ,$  and $g_3$ are respectively the U(1) , SU(2) and SU(3) gauge 
  couplings . Defining $v$ to be
\begin{equation}     
	v= (2 \sqrt{2}*G_F)^{-1/2}                                   
\end{equation}
	
   the on-shell masses of the top and Higgs are related to $v$ by
	\begin{equation}
	      m_t^2   = G(t= log(\frac{m_t}{m_Z}))* v^2                    
        \end{equation}
\begin{equation}
           m_H^2=   2*\lambda(t=log (\frac{m_H}{m_Z}))*v^2             
\end{equation}

  Eq.(\ref{two})which gives the running of the top coupling is
  homogeneous and is easily solved. I use 
\begin{equation}   
  [\alpha]^{-1} = 127.9 ;\quad  Sin^2 \theta_W = 0.231 ;\quad  and \quad \alpha_3
=0.119 
\end{equation}

	at t=0 and take the on-shell value of G(t) to be unity
 which corresponds to $m_t = v$ cf.eq.(5). Solving eq(2) with these
 inputs , and inserting the resulting G(t) in eqn (3) , the latter
 is solved numerically [8]in the range $ 0 < t <10$ using a range of
 values of $\lambda (0)$ as the initial value at t=0. 

\begin{figure}[t!]
\parbox{2.6in}{
\includegraphics[scale=0.4]{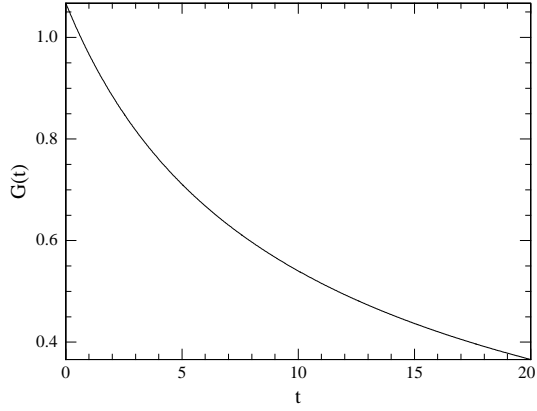}
%\special{psfile=fig1.eps hoffset=-20 voffset=-300 hscale=50 vscale=50}
%\vspace{2.4in}
}
\caption{The solution G(t) obtained from solving eq.(2).}
\end{figure}

\begin{figure}[t!]
\parbox{2.6in}{
\includegraphics[scale=0.4]{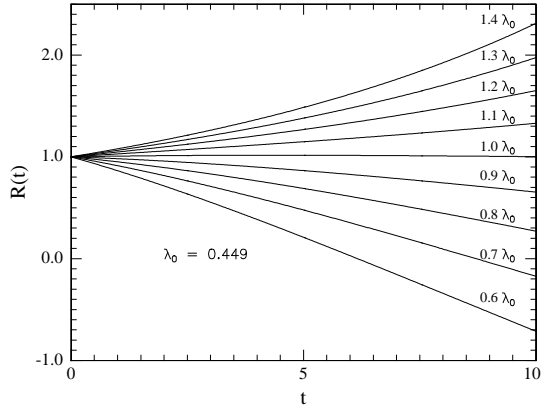}
%\special{psfile=fig2.eps hoffset=-20 voffset=-300 hscale=50 vscale=50}
%\vspace{2.4in}
}
\caption{The ratio $R(t)=\frac{\lambda(t)}{G(t)}$ corresponding to different 
initial values of $\lambda(0)$ expressed in units of $\lambda_0$.}
\end{figure}

 In Fig. 1 , the solution of Eq.(2) giving the square of the
 top coupling G(t) is plotted. It is  a decreasing function and
 drops from a value of 1.068 at t=0 to 0.711 at t=5 and further
 to 0.540 at t=10 . To find which value of $\lambda(0)$ will make
 the solution $\lambda(t)$ have the same  t dependence as G(t)
 I consider the ratio 

       $$ R(t) = \frac{\lambda(t)}{G(t)} * \frac{G(0)}{\lambda(0)}$$

  which is plotted in Fig.2. It is seen that the flow of R(t) is
  sensitively dependent on $\lambda(0)$. To find the best value of
  $\lambda(0)$ which satisfies the scale independence assumption 
   we determine  for different  initial values of $ \lambda(0)$ the values 
  $\Delta ( t_0 , \lambda(0) )$ which is 
  defined as the difference between the maximum and minimum  value 
  of R(t) in the interval $ 0 < t < t_0 $. This is displayed in
  Fig.3 for the two intervals  $t_0 = 5$ and 10.
  We take the value of $\lambda(0)$ corresponding to the minimum
  of $\Delta (t_0 , \lambda (0) )$ , to be the best estimate for
  $\lambda(0)$ and denote it by $\lambda_0$. If we use the interval
  upto $t_0= 10$ , we get the value 0.449 and if we reduce the search
  interval down to $t_0$ =5 we get 0.444 . Returning to fig.2 
  we see that there is a clear separation of trajectories R(t) 
  and even a ten percent departure in the value of $\lambda(0)$
  from $\lambda_0$ distinctly fails to meet the criterion of almost
  constant behaviour for R(t) . The origin of the various
  lower and upper bounds on the Higgs mass as a function of $\mu$
  i.e. the energy scale upto which we like the theory to be valid
  can be seen to follow from how far out in t one wants  to go.

\begin{figure}[t!]
\parbox{2.6in}{
\includegraphics[scale=0.4]{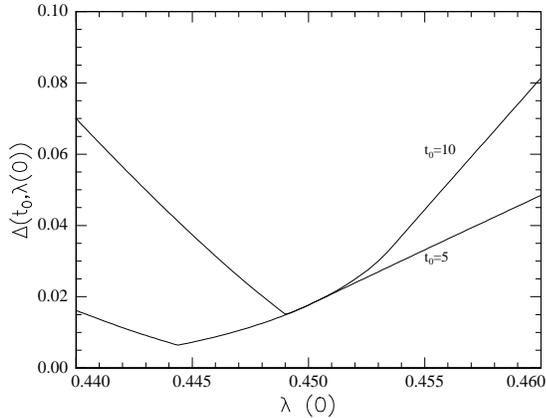}
%\special{psfile=fig3.eps hoffset=-20 voffset=-300 hscale=50 vscale=50}

}
\caption{The variation of $\Delta(t_0$,$\lambda_0)$ with $\lambda(0)$ for 
$t_0=10$ and $t_0=5$.}
\end{figure}

   From the optimal value $\lambda_0$ , we can use the mass-shell
   condition eq.(6) to obtain the Higgs mass. Using the
   value 0.449 at t=0 corresponds to the mass value
  
         $$ m_H  = 160 GeV  $$
  
   Since the Higgs mass has only square-root dependence
   on $\lambda$ , even a ten percent error in determining
   $\lambda_0$ is accurate enough to fix $m_H$ within say 10 GeV
   of the value given above . 
 
   This is  consistent with bounds obtained from an analysis 
  of precision data [ 9] and is quite close to the estimate of
  D'Agostini and Degrassi [10] who using  constrained analysis of direct 
  search and precision measurement measurements find an expected 
 value of 160-170 Gev for the Higgs mass as the central value with 
 a standard deviation of 50 to 60 GeV.

  We can find an estimate of $\lambda_0$ without having to solve
  eq.(3) as follows. If we assume R(t) is a constant then we can
  set its  derivative to be equal to zero. Implementing it at t=0
  we have  
\begin{equation}
   \lambda(0)* \frac{d}{dt} G(t)|_{t=0}      =       G(0) * \frac{d}{dt} \lambda (t)|_{t=0}     
\end{equation}
  With the use of eqns (2) and (3) this reduces to a quadratic
  equation for $\lambda(0)$ in terms of G(0) and the gauge couplings[11].
  It is instructive to consider three cases . $ a) g_1 = g_2  = g_3 =0$\quad
 $b) g_1=g_2  = 0$\quad c) all have their experimental values.

     Case a) $ \lambda(0) = G(0)\frac{\sqrt{17} -1/2 }{4}$ . If we had adopted 
     $m_t$ instead of $m_Z$ in defining t in eq.(1) , G(0) is the on shell value.
     Using eq.(6) then one gets $m_H/m_t$ = 1.36 . Case b) Using
	G(0)= 1.068 as noted earlier and eq.(6) one gets 
    $\lambda(0)$= 0.425 which indicates the important role of $g_3$. Case (c)
     One gets $\lambda(0)$ = 0.433 which  is close to the optimal
     values of $\lambda_0 $ obtained from fig.3. The discrepancy is not just 
    a matter of numerics. Although R(t) remains close to unity over a
    wide range it is not a constant. Computing the derivative 
    $\frac{d}{dt}R(t)$ using the numerical solution to eq.(3) we check that
    it is zero at $\lambda(0)$ =0.433 and has a non-zero but tiny
    slope of value = 0.0035 if $\lambda(0)$ = 0.449 and 0.0016
    if $\lambda(0)$ = 0.444 , which also explains the shift in
    $\lambda_0$ for the two cases considered in Fig.3. These 
    little variations can easily be changed by shifting the 
    t=0 point from $m_Z$.

	 It would be interesting to study the effect of
	 higher order terms in the RG equations
	which are important at lower energies and see whether
	 one can bootstrap the values of $G(0)$ and
	$\lambda (0)$.  
	
        {\bf Acknowledgements}: I am grateful to V.Srinivasan for
	valuable discussions on quasi-fermion boundstates in
         BCS type models and T.Ferbel about experimental
	program to discover the Higgs. I thank the High
	 energy group at Syracuse for their support and 
	helpful discussions.   
        I am very grateful to Abhay Pasupathy for his generous
        help with numerical computation and to him and Ankur
        Mathur for their help in preparing the manuscript.	
\vspace{0.1in}

	{\bf REFERENCES}
\begin{enumerate}
\item
	Y.Nambu and G.Jona-Lasinio , Phys.Rev.{\bf 122} (1961) 345 
        This paper is reprinted in the volume containing 
	selected papers of Nambu -" Broken Symmetry " Edited
	by T.Eguchi and K.Nishijima  World Scientific (1995)
	where one can find some of Nambu's work on BCS type theories extending 
	over condensed matter, nuclear and particle physics.
	 The 0:1:2 mass relation between the phase mode , fermion and 
	amplitude mode was also found by L.Leplae , F.Mancini
	and V.Srinivasan UWM-4867-71-12 unpublished.
   
\item
	W.A.Bardeen , C.Hill and M.Lindner,1990 ,Phys.Rev.D
	{\bf 41} , 1647. ; Y.Nambu in New Theories in Physics , Proceedings
	of the XI International Symposium on Elementary Particle
	 Physics, Kazmimierz ,Poland 1988 
	(1988 ) which is reprinted in Broken Symmetry cf.Ref [1]
	V.A. Miransky, M.Tanabashi , and K.Yamawaki ,Phys.Lett.B
	{\bf 221} , 177 (1989). A review of top-condensate models
	 containing several later
	developments can be found in G.Cvetic , Rev.Mod.Phys. 
	{\bf 71}(1999) 513 . 

\item N.Cabibbo, L.Maiani , G.Parisi and R.Petronzio ,
          Nucl.Phys. B{\bf 158} (1979) 295 
	   
\item M.A.B.Beg , C.Panagiotakapoulos and A.Sirlin ,
	Phys.Rev.Lett. {\bf 52} (1984) 883 . 

\item M.Lindner , Z.Phys. C{\bf 31} (1986) 295

\item A discussion of various bounds can be found
         in , C.Quigg Acta.Phys.Polon. B{\bf 30} 2145 (1999)

\item The normalisations of the couplings used here 
	 is same as used by Lindner[5]

\item  After solving eq.(1) for G(t) , Lindner [5] expresses 
	the integrating factor in terms of simple functions
        which is an excellent approximation.  This
	is very useful to speed up  the numerical solution of eq.(3).
	I have adapted his procedure , taking into account the
	small difference in gauge couplings used here as compared
        to his values. 

\item W.J.Marciano , hep-ph/0003181 ;  A.Sirlin
	  hep-ph/9912227; C.Quigg , hep-ph/0001145

\item G.D'Agostini and G.Degrassi hep-ph/0001269

\item The procedure here is similar to the work of
         J.Kubo, K.Sibold and W.Zimmerman, Nucl.Phys. B{\bf 259},
	 (1985) 331 ; Phys.Lett B{\bf 220} 185 (1989) who use the 
	 reduction of coupling constants due to W.Zimmerman,
	 Commun.Math.Phys. {\bf 97} (1985) 211 : R.Oehme and W.Zimmerman
	 Commun.Math.Phys. {\bf 97} (1985) 569.  There are some differences.
	 First I do not attempt to relate the gauge couplings 
	 to the Yukawa
         coupling as these authors do , secondly the 
         simplified procedure of eqn(8) is used after having verified the
         near constancy of R(t) as otherwise setting the 
         derivative equal to zero at some t value would
	be without justification.
 
\end{enumerate}          
          
\end{document}